\documentclass[superscriptaddress,showpacs,twocolumn,floatfix,amsmath,footinbib,amssymb]{revtex4-1}
\usepackage{amssymb}
\usepackage{amsmath}
\usepackage{epsfig}
\usepackage{t1enc}
\usepackage{soul}
\usepackage{color}
\usepackage{verbatim}
\usepackage{graphicx}
\usepackage{layout}
\usepackage{times,txfonts}
\newcommand{\beq}{\begin{equation}}
\newcommand{\eeq}{\end{equation}}
\newcommand{\bea}{\begin{eqnarray}}
\newcommand{\eea}{\end{eqnarray}}

\begin{document} 
\raggedbottom
\title{Ruling out stray thermal radiation in analogue black holes}
\author{Jason Doukas}
\affiliation{School of Mathematical Sciences, University of Nottingham, Nottingham NG7 2RD, United Kingdom}
\author{Gerardo Adesso}
\affiliation{School of Mathematical Sciences, University of Nottingham, Nottingham NG7 2RD, United Kingdom}
\author{Ivette Fuentes}
\affiliation{School of Mathematical Sciences, University of Nottingham, Nottingham NG7 2RD, United Kingdom}
\date{\today}
\begin{abstract}
Experimental searches for the thermal radiation from analogue black holes require the measurement of very low temperatures in regimes where other thermal noises may interfere or even mimic the sought-after effect. In this letter, we parameterize the family of bosonic thermal channels which give rise to such thermal effects and show that by use of coherent states and homodyne detection one can rule out the non-Hawking contributions and identify those candidate sources which arise from Hawking-like processes.
\end{abstract}
\maketitle
{\it Introduction}.---The ability to measure Hawking radiation in analogue black hole systems \cite{analogue, Barcelo2011} with high experimental control is approaching. Several experimental groups are showing signs of analogue black hole production and there is cautious optimism that soon we will have direct measurements of analogue Hawking radiation and be able to perform quantum precision tests of this remarkable theory \cite{Faccio2013}.

While there is evidence for Hawking radiation from the stimulated emission of white holes in the scattering of shallow water (gravity) waves \cite{Silke, Unruh2014}, thermal fluctuations in water at room temperature limit further progress in this direction. On the other hand experiments with ultrashort laser pulse filaments \cite{Belgiorno} and Bose-Einstein Condensates (BECs) \cite{Lahav2010} provide the opportunity to go one step further and investigate the quantum aspects of this effect.

To measure the Hawking radiation one needs to observe thermal radiation at a temperature proportional to the analogue surface gravity (the derivative of the flow velocity at the horizon).  However, at low temperatures there are potentially many sources of stray thermal radiation. For example, in the case of analogue black holes that exploit the non-linear Kerr effect, the intensity of the pulse is required to be so high that it begins to damage the medium in which it propagates thus introducing broadband noise  \cite{schutzhold2011, Unruh2014}. The possibility exists that these other sources of thermal noise can in some ways mask, or give false signatures for, the actual Hawking effect. An important and timely demand for concrete implementations of analogue gravity is therefore to develop additional criteria to experimentally determine whether or not any observed radiation does in fact originate from a Hawking-like process. 

In this letter we investigate, using the general formalism of bosonic channels and adapting methods from quantum estimation theory \cite{parameest}, what kind of processes could mimic the Hawking effect and how these impostor scenarios can be ruled out with quantum experiments. This adaptation of quantum parameter estimation to solve a timely experimental problem in quantum field theory in curved space-times is preceded by several pioneering works on this subject in different contexts, see for example \cite{Aspachs, Downes, ivettepapers, Doukas2014}.

Since we consider the detection of thermal radiation to be a necessary ingredient, let us assume that an experimental group has demonstrated a candidate thermal effect in the sub-luminal region. We ask the question, what other processes could possibly be responsible for this thermal radiation from the vacuum? One could approach this by investigating all the possible physical processes that could operate in each given experiment. However, such processes will be highly dependent on the implementation of the analogue system. One would then be faced with the problem of investigating all the potential non-Hawking thermal processes for each system individually. In this work, we analyze an abstract parametrization of all possible processes compatible with the observed thermality, without the need to specialize to a particular physical setup. We then show that quantum precision measurements can be used to experimentally constrain the abstract parameter space (giving rise to thermal effects), and to rule out the potential non-Hawking processes.

There are many ways an experimental group may try to rule out stray thermal sources. The most direct one would be by detecting quantum correlations between particles created at the analogue horizon. In this letter we will assume that the experimenter, perhaps because of the experimental difficulty of their measurement, or due to intrinsic limitations, cannot access the correlations either classical \cite{ccorrelations} or quantum \cite{qcorrelations} present between the super-luminal and sub-luminal regions. We instead focus on what can be learned from the observations in the sub-luminal region alone. The extension to methods encompassing the detection of correlations will be the subject of a future work \cite{futurework}.

Condensed matter black hole analogues have an acoustic metric that is conformal to the Schwarzschild black hole metric  \cite{Visser1998, Barcelo2011} in Painlev\'{e}-Gullstrand-Lema\^{i}tre coordinates \cite{Painleve}. Up to conformal factors, the line-element in these coordinates ($\hbar=c=k_B=1$) is:
\bea
ds^2=dt_p^2-[dx+v(x)dt_p]^2,
\eea
where $v(x)$ is the fluid flow velocity in the negative $x$ direction. The black hole horizon occurs when the fluid flow equals the speed of sound $v=1$. In analogue systems cutoff scale effects give rise to a modified wave equation:
\bea
\square\phi=D(\partial_x)\phi,
\eea
with the effect of modifying the dispersion relation at high wave-numbers. For example, in BECs $D=-\gamma^2\partial_x^4$, where $\gamma$ is related to the healing length. In Einsteinian gravity similar modifications are speculated to arise from Planck scale corrections \cite{Transplankian} to the theory.

The picture of how the emission occurs in this situation is quite different to the one originally conceived by Hawking \cite{Hawking1975}. The wave-packet that eventually gives rise to the Hawking particles originates from either inside the horizon (for super-luminal dispersion at large wave-numbers) or outside (sub-luminal dispersion at large wave-numbers).

It was shown in \cite{Schuetzhold2013} that if the ultraviolet cut-off scale is much larger than the surface gravity $1/\gamma\gg\kappa$ then the WKB method is valid (in appropriately chosen coordinates) throughout the entire evolution.  The initial wave-packet is assumed to be a superposition of positive frequency waves with respect to the Kruskal null coordinate,  $U=-\kappa^{-1}\text{exp}(-\kappa t_p-\kappa\int\frac{dr}{v(r)-1})$. Canonical quantization then associates an annihilation operator $\hat{A}_{\Omega}$ to the initial wave-packet, where $\Omega\equiv2\pi\omega/\kappa$ and $\omega$ is the ordinary time frequency of the wave-packet when it is far from the horizon.  As this wave-packet evolves towards the horizon it is ripped into two wave-packets traveling in opposite directions. The late time wave-packet on the sub-luminal side is associated with the annihilation operator $\hat{b}^{\dagger}_{\text{I}}$, whereas the late time wave-packet on the super-luminal side is associated with the annihilation operator $\hat{b}^{\dagger}_{\text{II}}$. Quite remarkably, the thermal emission is extremely robust even in the presence of the modified dispersion relation and the transformation can be described by a two-mode squeezing operation:
\bea\label{tms}
\hat{A}_{\Omega}=\cosh{r_{\Omega}} \hat{b}_{\text{I}\Omega}+\sinh{r_{\Omega}} \hat{b}^{\dagger}_{\text{II}\Omega},
\eea
where $r_{\Omega}=\text{arctanh}(e^{-\Omega})$,   

The process transforms the initial state of the impinging wave-packet mode into a (thermalized) state in the black hole exterior. This final state is found by performing the two-mode squeezing operation (\ref{tms}) and tracing over the mode $\hat{b}_{\text{II}}$ that is swept inside the black hole. The transformation that takes the initial state to the thermalized output state is a non-unitary single mode map known as bosonic amplification channel \cite{Aspachs, amplificationchannel}.

Any bosonic channel can be represented in terms of two real $2n\times 2n$ matrices $(X,Y)$, where $n$ is the number of input modes of the channel. Given an input Gaussian state with first moments, ${\bf d}$, and covariance matrix (second moments), ${\bf \sigma}$, the bosonic channel transforms the state according to:
\bea
\mathbf{ d}'&=&X \mathbf{ d} ,\\
\boldsymbol{\sigma}'&=&X\boldsymbol{\sigma} X^{T} +Y.
\eea
Single mode bosonic  channels that take thermal states (including the vacuum) to thermal states are called single mode thermal channels \cite{Schafer}. In this case, the matrices $X$ and $Y$ can be expressed as:
\bea\label{matrixX}
X=\left(\begin{array}{cc} \sqrt{|x|} & 0\\ 0 & \text{sgn}(x)\sqrt{ |x|}\end{array}\right);\quad \text{and } Y=\left(\begin{array}{cc} y & 0\\ 0 & y\end{array}\right),
\eea
where physical channels must satisfy $y\geq|x-1|/2$. 

\begin{figure}
\centering
\includegraphics[width=8cm]{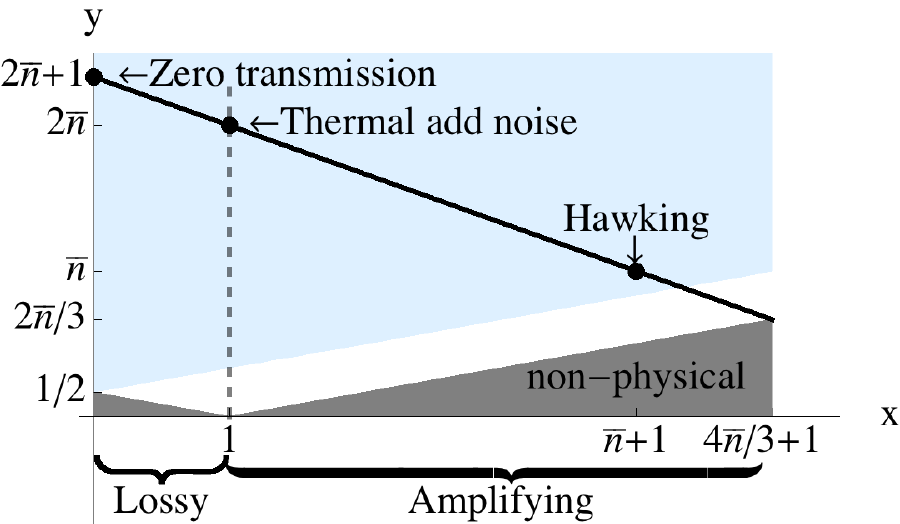}
\caption{\label{fig:thermalchannel} (Color online) We plot the relevant part of the parameter space of thermal channels parameterized by $(x,y)$ defined in equation (\ref{matrixX}). The thick black line corresponds to thermal channels taking the vacuum state to the thermal state with mean occupation number $\overline{n}$. The Hawking channel is represented by one point on this line at $y=\overline{n}$. Also shown is the classical thermal add noise channel and the zero transmission thermal channel (see text). The shaded grey region is an unphysical part of the parameter space and the shaded light blue region indicates that the channel is \textit{entanglement breaking}, meaning that any entanglement initially present between the state and an ancilla state (not affected by the channel) is destroyed once the initial state has passed through the channel.}
\end{figure}

The covariance matrix of a thermal state has the diagonal form $\boldsymbol{\sigma}'=(2\overline{n}+1)I_2$ where $I_2$ is the $2\times2$ identity matrix and $\overline{n}$ is the average thermal occupation number of the mode. For a quasi-monochromatic wave-packet mode of peak frequency $\omega$, the mean thermal number is related to the temperature, $T$, by $\overline{n}=(e^{\omega/T}-1)^{-1}$. Therefore, if one prepares the vacuum state and measures a thermal state of mean particle number $\overline{n}$ one arrives at a constraint on the channel parameters given by:
\bea\label{parameterrelation}
2\overline{n}+1=|x|+y.
\eea
If we assume that $\overline{n}$, is measured to very high accuracy we can use this relation to eliminate one of the channel parameters. Therefore, all possible Gaussian channels that take the vacuum state to the observed thermal state with mean particle number $\overline{n}$ are characterized by a single real parameter $y$. For physically allowed channels the $y$-values are restricted from $2\overline{n}/3$ to $2\overline{n}+1$. A picture of the allowed channels is shown in Fig.~\ref{fig:thermalchannel}, where those thermal channels which take the vacuum state to the thermal state with mean occupation number $\overline{n}$ lie on a straight-line in this parameter space.

The Hawking channel (amplification channel) is given by the conditions that $x=\overline{n}+1$ and $y=\overline{n}$.  Another channel of interest is the classical add noise channel in which $y=2\overline{n}$ and $x=1$. This would correspond to thermal radiation being added to the initial state radiating from the direction of the horizon, we could imagine such low energy radiation arising from random stochastic noise. Another channel to be distinguished from the classical add noise channel is the zero-transmission channel in which $x=0$ and $y=2\overline{n}+1$.  This channel corresponds to the situation in which all the initial information is totally erased and replaced with thermal radiation at the output. Of course, there is a whole continuous parameter range of possible channels in between those just described. Those with $x<1$ are lossy channels, while those with $x>1$ increase the thermal occupation number of the input state.

We see that there are many ways in which a thermal state can be obtained from the vacuum. Our objective is to somehow learn what channel is actually operating in a given experiment and in particular whether or not the channel is of the Hawking type. Given an initial vacuum state there is clearly no measurement to distinguish the channels, since all give the same thermal output state by definition. However, if we could prepare different initial states, then the output would no longer be purely thermal, indeed it would change in a way that was dependent on the channel. By performing measurements on the output state, one could distinguish the different possible values of $y$ from the measurement outcomes. To determine the best strategy to measure the thermal channel we will use the theory of quantum parameter estimation \cite{parameest}.

In quantum parameter estimation, given some input state, $\rho_0$, one searches for the best measurement strategy to determine the value of a not directly observable (continuous) parameter associated to a transformation of the input. In our case the output state is $\rho_y$, where $y$ indicates that the initial state was acted on by the thermal channel corresponding to $y$. Using a general positive operator-valued measurement $\{\hat{O}_\lambda\}$, one obtains a probability distribution $p(\lambda|y)=\text{Tr}[\hat{O}_\lambda\rho_y]$ from the measurement outcomes.  For $N$ repetitions of the experiment, the variance of the parameter is bounded by the Cramer-Rao inequality \cite{Cramer, Helstrom, Holevo}, $(\Delta y)^2\geq1/NF(y)$, where
\bea\label{FI}
F(y)\equiv\int p(\lambda|y)\left(\frac{d \log {p(\lambda|y)}}{dy}\right)^2d\lambda,
\eea
is the Fisher information. The Fisher information is bounded above by the Quantum Fisher Information (QFI) \cite{Braunstein}, $H_y$ which can be written as:
\bea
H_y(\rho_y)=\lim_{dy\rightarrow0}(1-4\mathcal{F}\big(\rho_y,\rho_{y+dy}\big)/dy^2,
\eea
where $\mathcal{F}(\rho_1,\rho_2)\equiv\text{Tr}[\sqrt{\sqrt{\rho_1}\rho_2\sqrt{\rho_1}}]^2$ is the fidelity. Note that the QFI evaluates the fidelity between two infinitesimally separated output states near the channel $y$ and corresponds to the Fisher information of a measurement which saturates the Cramer-Rao bound. The QFI thereby provides a measure of the ultimate precision attainable for a given probe state. By fixing the energy of the probe state, one can compare the effectiveness of different probe states in revealing the channel parameter $y$.

If the Hawking process was the dominant thermal process one would expect to obtain a value of $y\sim \overline{n}$. In order to perform this type of test the experimentalist needs to prepare the different initial states in the specified mode. Since we do not want to limit our analysis to any particular experimental setup, for simplicity we will assume that any Gaussian operation is available and restrict the class of initial states that we consider to coherent states, squeezed states and thermal states. The analysis can then be repeated for those states which are available when one considers a specific experimental setup. Although there may be experimental challenges involved in preparing such generic states these would not seem to be insurmountable for any fundamental reason. Indeed, the possibility of preparing such states in photon systems, like that of ultrashort laser pulses, appear viable and recent progress in the squeezing of BEC phonons \cite{Jaskula2012} is also very promising.

\begin{figure}
\centering
\includegraphics[width=8cm]{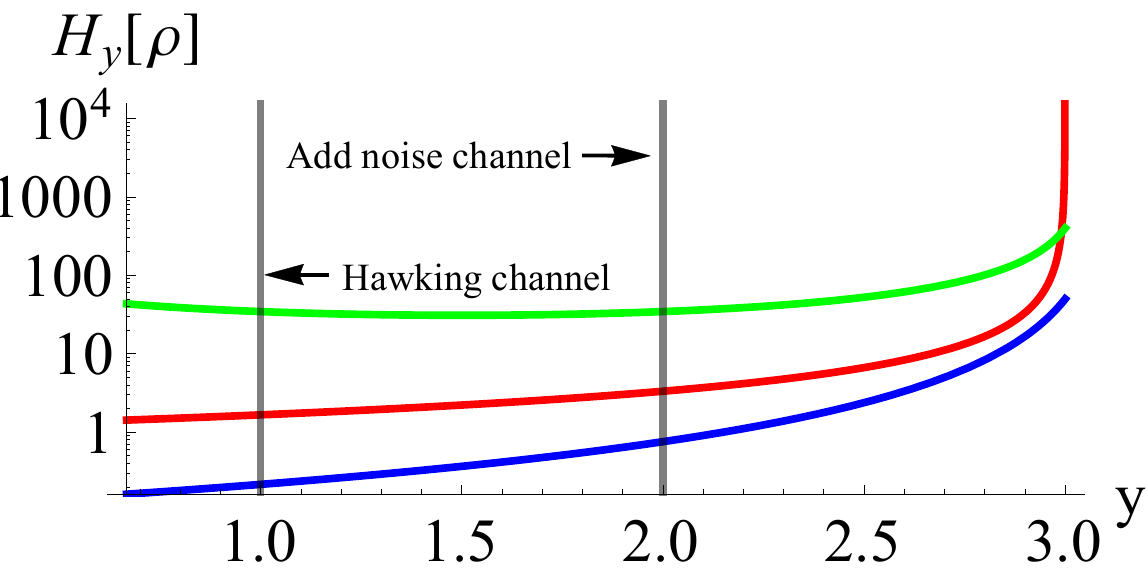}
\caption{\label{fig:cfstrategies} (Color online) Quantum Fisher information for three initial states: (top green) squeezed state; (middle red) coherent state; (bottom blue) thermal state. The plot is for the specific choice of parameters $\overline{n}=1$ (average vacuum--thermal number) and $n_0=10$ (mean initial particle number). Grey vertical lines indicate the position of the Hawking channel ($y=1$) and the classical add noise channel $y=2$. While the shape is slightly modified we find a similar ordering of the strategies for (realistic) low temperature of horizon radiation $\overline{n}\ll1$.  }
\end{figure}

{\it Results.}---In Fig.~\ref{fig:cfstrategies} we compare the QFI that could theoretically be attained for the different choices of initial states. We consider coherent states of mean particle number $n_0$, squeezed states with $\sigma=\text{diag}\{e^{2s},e^{-2s}\}$ where $n_0=\sinh{s}^2$, and thermal states of mean particle number $n_0$. We find that squeezed states always give the greatest information in the parameter space of interest, namely near $y\sim \overline{n}$. In the limit of large initial energy, $n_0$, we find the simple expressions for the QFI:
\bea
H_y(\rho_{\text{coherent}})&=&\frac{n_0}{(1+2\overline{n})(1+2 \overline{n} -y)},\\
H_y(\rho_{\text{thermal}})&=&\frac{1}{(1+2\overline{n}-y)^2}+\mathcal{O}(n_0^{-1}),\\
H_y(\rho_{\text{squeezed}})&=&\frac{3}{4}\frac{(1+2\overline{n})^2n_0}{y(1+2\overline{n}-y)}+\mathcal{O}(n_0^{-1}).
\eea
Note that the QFI for the coherent state is exact.

In all cases the QFI depends on the  measured parameter, $y$. Since the QFI of the coherent state scales linearly with the initial energy the coherent state makes a very practical probe state. I.e., we can get as much information as is required simply by pumping up the intensity. While squeezed states also have this behavior at large energies, coherent states are in general easier to prepare and are therefore preferred.

The extent to which the optimal performance can be achieved will depend on the technological limitations that are specific to the analogue spacetime implementation. In general, the measurement which attains the QFI may be quite difficult to perform in practice \cite{Braunstein, Monras}. However, we find that for a coherent state with homodyne measurements the Fisher information is maximized by measuring in the direction of displacement and is given by $F(y)=H_y(\rho_{\text{coherent}})$.
In other words, the QFI can be attained for coherent state probes (at any energy) simply by performing homodyne measurements. Such measurements are readily done in optics experiments and would appear to be implementable in analogue experiments involving  photons. In the phonon case, the reference field could be created by Bragg scattering techniques \cite{Stamper}.

One can verify that a coherent state and homodyne measurements should provide a good strategy to experimentally determine the channel parameter $y$ by the following simple argument. The coherent state provides gain such that the quadrature measurement (in the direction of the displacement) yields \cite{footnote1}:
\bea\label{cohxquad}
\langle \hat{q}\rangle= 2\sqrt{2\overline{n}+1-y}\sqrt{n_0}.
\eea
Thus $y$ can be determined from the experimentally obtained values of $\langle \hat{q}\rangle$ and $\overline{n}$ and the known intensity of the coherent state $n_0$. Furthermore, since $\Delta q=2\overline{n}+1$, the relative uncertainty of the measurement $\Delta q/q\sim 1/\sqrt{n_0}$ decreases like the inverse square root of the initial energy. Therefore, the statistical certainty can always be improved by increasing the intensity of the initial coherent probe state. 

So far, we have assumed that the experimentalist can prepare an initial vacuum state in order to measure the temperature of the candidate Hawking radiation and subsequently determine $\overline{n}$. Obviously this is only an approximation since one can never go to absolute zero temperature. For example, in non-linear Kerr media there is an ambient temperature of the environment that the fibre is contained within, and even in ultra cool systems like BECs the BEC itself will necessarily have some small but non-zero temperature. One might think such situations can already be accommodated for in our setup by considering initial probe states that are very weakly thermalized at this ambient temperature, $n_T$. The issue is that the method we have adopted required measurement of $\overline{n}$ to eliminate one of the channel parameters, namely $x$. However, if the initial state of the field is not at zero temperature, then the modified measured value of thermal radiation from the horizon will carry over into an incorrect identification of $x$ which introduces an error in the channel identification. 

Fortunately, we can handle this situation even when we don't know a priori what is the actual thermal channel. Experiments using the $n_T$ thermal state will yield a modified temperature with mean particle number:
\bea\label{modnumber}
2n'+1= (2\overline{n}+1-y)(2n_T+1)+y,
\eea
where $\overline{n}$ is still defined to be the thermal number associated with the action of the channel on the true vacuum state.  This error can be corrected if one knows the ambient temperature of the system $n_T$ and the measured value $n'$ of radiation from the horizon. Solving for $\overline{n}$ we obtain:
 \bea
2\overline{n}+1=\frac{2n'+1-y}{2n_T+1}+y,
 \eea
 in which case the channel is given by (\ref{matrixX}) with the same free parameter $y$ but now $x$ takes the value:
 \bea
 x=\frac{2n'+1-y}{2n_T+1}.
 \eea

 A simple calculation reveals that preparation of a coherent state of mean photon number $n_0$, will lead to gain in the $\hat{q}$ quadrature according to:
\bea
\langle \hat{q}\rangle= 2\sqrt{\frac{2n'+1-y}{2n_T+1}}\sqrt{n_0},
\eea
which reduces to equation (\ref{cohxquad}) when $n_T\rightarrow0$. Therefore, the measured quadrature value $\langle \hat{q}\rangle$, the observed temperature of thermal radiation from a horizon in the pseudo vacuum state, $n'$,  and the ambient temperature of the BEC under normal (no horizon) conditions, $n_T$, allow one to determine the channel value $y$. With these modifications one would expect that even with non-zero initial temperature, one would obtain $y\sim \overline{n}$ for Hawking effects at the horizon.

{\it Conclusion.}---By parameterizing the set of thermal channels and applying tools from parameter estimation we have formulated quantum tests that could be used to rule out non-Hawking effects in analogue black hole experiments. The real benefit of this approach is that one does not need to have a complete understanding of the fundamental processes of the system in order to rule out the alternatives. Thereby offering the use of these techniques in any analogue black hole setup.

We found the nice strategy of coherent state preparation and homodyne measurements can be used to determine the thermal channel parameter $y$, and the uncertainty in this parameter could be made arbitrarily small simply by turning up the intensity of the coherent state. At low energies one could obtain even better performance by using squeezed states. However, in practice setups involving squeezed states are likely to be more difficult to engineer. 

The coherent state strategy resembles the strategy recently suggested to discriminate the Unruh effect from non-amplifying theories \cite{Doukas2013}. However, in that case a Kennedy receiver was proposed to implement the measurement which is considerably more difficult than homodyne detection. Measuring the Hawking thermal channel with coherent states, as described in this letter, is extremely promising in light of the fact that homodyne measurements are so easy to implement. 

We first presented results for the ideal case in which the system could be cooled to arbitrarily small temperatures. We then showed that even non-zero initial temperatures could be accounted for within our analysis. Our main conclusion is that if the ambient thermal temperature is properly accounted for, then one would observe a value of $y$ equal to the the mean thermal number of the Hawking thermal state at a temperature proportional to the analogue surface gravity. Estimation of the channel parameter near the expected value $y\sim \overline{n}$ would certainly rule out a large class of alternatives and give a lot of support to the interpretation of the temperature as a Hawking effect. 
In this work we did not utilize the additional information that exists in the correlations between the inside and outside of the black hole, which may be inaccessible in certain analogue implementations. This information will provide even further clues into the nature of the processes taking place and would be a very interesting path to pursue in future.

The authors thank J. Steinhauer for valuable discussions and for bring to our attention Bragg diffraction techniques. J.D. thanks M. Ahmadi and A. Lee for useful discussions. Financial support from  EPSRC (CAF Grant No. EP/G00496X/2 to I. F.) and FQXi (Grant No. FQXi-RFP3-1317 to G. A.) is acknowledged.


\begin{thebibliography}{99}
\bibitem{footnote1}
We use the convention that $\hat{q}\equiv\hat{b}_{I}+\hat{b}_{I}^{\dagger}$, and that the covariance matrix of the vacuum is normalized to unity.
\end{thebibliography}
\end{document}